\documentclass[sn-basic]{sn-jnl}

\theoremstyle{thmstyleone}%
\theoremstyle{thmstyletwo}%
\theoremstyle{thmstylethree}%
\usepackage{appendix}
\usepackage[english]{babel}
\usepackage{amsmath, amsfonts, amssymb, mathrsfs}
\usepackage{natbib}
\usepackage{subfigure}

\begin{document}

\title [\hspace{1 pt}]{ Pesticide Mediated Critical Transition in Plant-Pollinator Networks }

\author[1]  {\fnm{Arnab} \sur{Chattopadhyay}} 

\author[1]  {\fnm{Amit} \sur{Samadder}} 

\author*[1]{\fnm{Sabyasachi} \sur{Bhattacharya}} \email{sabyasachi@isical.ac.in}

\affil[1]{\orgdiv{Agricultural and Ecological Research Unit}, \orgname{Indian Statistical Institute}, \orgaddress{\street{203, B. T. Road}, \city{Kolkata}, \postcode{700108}, \state{West Bengal}, \country{India}}}

\abstract{ Mutually beneficial interactions between plant and pollinators play an essential role in the biodiversity, stability of the ecosystem and crop production. Despite their immense importance, rapid decline events of pollinators are common worldwide in past decades. Excessive use of chemical pesticides is one of the most important threat to pollination in the current era of anthropogenic changes.  Pesticides are applied to the plants to increase their growth by killing harmful pests and pollinators accumulates toxic pesticides from the interacting plants directly from the nectar and pollen. This has a significant adverse effect on the pollinator growth and the mutualism which in turn can cause an abrupt collapse of the community however predicting the fate of such community dynamics remains a blur under the alarming rise in the dependency of chemical pesticides.  We mathematically modeled the influence of pesticides in a multispecies mutualistic community and used 105 real plant-pollinator networks sampled worldwide as well as simulated networks, to assess its detrimental effect on the plant-pollinator mutualistic networks. Our results indicate that the persistence of the community is strongly influenced by the level of pesticide and catastrophic and irreversible community collapse may occur due to pesticide. Furthermore, a species rich, highly nested community with low connectance and modularity has greater potential to function under the influence of pesticide.  We finally proposed a realistic intervention strategy which involves the management of the pesticide level of one targeted plant from the community. We show that our intervention strategy can significantly delay the collapse of the community. Overall our study can be considered as the first attempt to understand the consequences of the chemical pesticide on a plant-pollinator mutualistic community. }


\maketitle

\section{Introduction}
\label{introduction}

Plant-pollinator interaction is vitally important to terrestrial ecosystems and to crop production. It plays a key role in plant community assembly \citep{bruno2003inclusion, ollerton2011many, wright2017overlooked, hale2020mutualism} and thus provides critical ecosystem services with immense economic and aesthetic value \citep{klein2007importance, potts2016safeguarding, requier2022bee, gallai2009economic}. Despite their immense importance, rapid decline events of many pollinators are frequent in recent years \citep{burkle2013plant, rhodes2018pollinator}.An important issue today is to determining how human activities impact the varied relationships between plants and their insect pollinators for the sake of conservation. 

Some major anthropogenic threats to pollinators in the face of current global changes includes habitat loss and fragmentation, chemical contamination, warming and climate change, parasite infection and invasion of alien species \citep{harrison2015urban, nicolson2017plant, potts2010global, dicks2021global}.  Habitat loss and fragmentation due to the transformation of grasslands into farmland or urbanization, reduces the fraction of interaction between the plant and their pollinators and hence decreases the persistence of the system \citep{mcwilliams2019stability, spiesman2013habitat}. Climate warming affects the pollination by altering the abundance and distribution of plants \citep{arft1999responses, inouye2003environmental} and by creating temporal mismatch between plant and pollinators \citep{hegland2009does}. Disease by parasite infection has caused some major pollinator decline events in recent years \citep{goulson2015bee}. Invasive species can reorganize the interactions, thus posing a risk to community stability\citep{vanbergen2018risks}. These several factors can cause an abrupt and often irreversible community collapse of such mutualistic network, known as tipping \citep{lever2014sudden, memmott2004tolerance}. Although there is a long list of potential causes of the decline of pollinators, the role of toxic contamination through agricultural pesticides and fertilizers, etc. is significant but rarely studied.

The intensification of agriculture and increasing reliance on agrochemicals makes pollinators  chronically exposed to contaminations. Contamination from agriculture includes commonly used pesticides \citep{sponsler2019pesticides}, fungicides and herbicides \citep{belsky2020effects}, and heavy metal contamination from the soil fertilizers \citep{nieminen2001effect}. Pollinators may be exposed to contamination in numerous ways, mainly thorough ingestion of contaminated pollen and nectar \citep{mitchell2017worldwide}, or exposure to contaminated nesting sited or materials which can impose variety of lethal and sub-lethal effect on them \citep{whitehorn2012neonicotinoid, henry2012common, tsvetkov2017chronic, cresswell2011meta, bryden2013chronic, godfray2014restatement}. Lethal or direct effect includes reduction in the growth rate \citep{whitehorn2012neonicotinoid} and rise in the mortality rate of pollinator and their larvae due to the accumulated contamination \citep{henry2012common, tsvetkov2017chronic}. Moreover, impairment of normal biological behaviours of pollinators such as memory and olfactory learning, navigation, foraging and feeding behaviour is seen due to contamination accumulation in several studies \citep{schneider2012rfid, sponsler2019pesticides, piiroinen2016chronic}. Furthermore, delayed larval and pupal developments is observed in laboratory studies which can decrease the time niche overlap between plant and pollinators for seasonal flowering plants \citep{wu2011sub}. Overall the factors diminish the strength of the plant-pollinator mutual disruption by altering the  pollinator visitation rate \citep{sponsler2019pesticides} and can be considered as sub-lethal effects. Exposure to chemical contamination can also compound the effects of other stressors on pollinator populations, such as loss of habitat and exposure to pathogens and diseases \citep{di2013neonicotinoid}.

Mathematical models played an important role to access the effect of current anthropogenic changes on plant-animal mutualistic community. Mutualistic networks are highly heterogeneous in degree distribution \citep{bascompte2007plant}, has moderate connectance \citep{valdovinos2019mutualistic}, moderate modularity \citep{olesen2007modularity} and importantly, high nestedness \citep{bascompte2003nested, burgos2007nestedness, zhang2011interaction}.  Nestedness promotes the species coexistence by reducing the interspecific competition \citep{bastolla2009architecture}, promoting complexity-stability relationship \citep{okuyama2008network}. These structural properties are responsible for the abrupt collapse of the whole community \citep{lever2014sudden}, i.e., the tipping, at some critical threshold of declining mutualistic strength or increasing mortality of pollinators, in the face of current anthropogenic changes. Metacommunity models of mutualistic species is studied considering habitat loss as a parameter, showing that there is a habitat loss threshold after which whole community collapses \citep{fortuna2006habitat} and the number of interaction of the network reduces suddenly \citep{fortuna2013habitat}. Other environmental stressors affecting mutualistic interaction strength by causing phenological change of species can magnify the effect of habitat loss, when acts together \citep{revilla2015robustness}. Epidemic models are developed \citep{truitt2019trait, proesmans2021pathways} to study the disease spread on plant-pollinator network, and nestedness is shown to promote disease persistence as highly generalist plants acts as hubs of pathogen transmission. A mathematical model incorporating the effect of the temperature on the psychological trait parameters such as growth, mortality etc. of a mutualistic network is studied \citep{bhandary2022rising}, where global temperature rise is shown to cause abrupt pollinator decline.

Environmental contamination, one of the detrimental consequences of current global changes, is an important driver of species demographic properties of such mutualistic communities. Few previous studies investigated the effect of toxins for antagonistic interactions \citep{huang2015impact, garay2013more}, but the understanding for mutualistic interaction remains a blur. \cite{wang2020persistence} recently studied a single plant-pollinator system under pesticide exposure. Transitions to the bistable state and consequently pollinator extinction emerges as a result of increasing pesticide use. In real world, a plant-pollinator mutualistic community involves multiple species and posses significantly different topological properties from the antagonistic networks which can largely influence its dynamics \citep{bastolla2009architecture, lever2014sudden} under the environmental stress. So study on the role of contamination on a multispecies mutualistic community is still lacking to best our knowledge. 

To address this gap, we considered a mathematical model of a multispecific mutualistic community, where plants are exposed to environmental contamination. Pollinators accumulates contamination from the interacting plants which alters several trait parameters of the system. Our research addressed following interrelated questions: (1) what is the essential role of toxin on the persistence and abrupt collapse of a mutualistic community? (2) what is the role of several network topological properties to maintain the diversity of the system on the onset of environmental contamination? This will be done with the help of a mathematical model of multispecies mutualistic community, where  physiological traits of pollinators are influenced by environmental contamination accumulated from the plant.  Furthermore, we proposed here a possible intervention strategy related to the network structure of the system to evade the tipping point of the community. 

Our paper is organized as follows. In the section \ref{method}, we proposed the mathematical model with the incorporation of contamination. We described our results in section \ref{results}. Finally our paper ends with a detailed discussion (section \ref{discussions}).

 \section{Method}
\label{method}

\subsection{Mathematical model }
\label{Mathematical model}
An ecologically realistic mathematical model of a mutualistic community incorporates the following basic properties such as intrinsic growth, intra and interspecific competition and mutualistic interaction between plant and pollinators. Let $P_{i}$ and $A_{i}$ be the abundance of the $i-th$ plant and pollinators, respectively. Following \cite{bastolla2009architecture}, the equations for the rate of change of  $P_{i}$ and $A_{i}$ are given by:

\begin{equation}
\label{network model with C}
\begin{split}
&\frac{dP_i}{dt}=P_i\bigg(\alpha^{P}_{i}-\sum_{j=1}^{S_{P}}{\beta^{P}_{ij}P_j}+  \sum_{j=1}^{S_{A}} m^{P}_{ij} \bigg) +u^{P}\\
&\frac{dA_i}{dt}=A_i\bigg(\alpha^{A}_{i}-\sum_{j=1}^{S_{A}}{\beta^{A}_{ij}A_j}+  \sum_{j=1}^{S_{P}} m^{A}_{ij} \bigg)- \kappa^{A} A_{i} +u^{A}
\end{split}
\end{equation}
, where $S_{P}$ and $S_{A}$ are the plant and pollinator richness in the community.  Description of the other parameters are the following. $\alpha^{P}$ and $\alpha^{A}$ are the intrinsic growth rate of plant and pollinators, respectively, in the absence of competition and mutualism.   The degree of the mutualism can be categorize in two ways, obligate and facultative, depending the sign of $\alpha$. If the population persists in the absence of mutualism, it is called facultative mutualism and $\alpha$ is positive in this case. On the opposite, $\alpha$ is negative for the case of obligate mutualism, where species cannot persists in the absence of mutualism. We assumed a common value $ \alpha $ as the intrinsic growth rate of all species, for the sake of simplicity. $\beta_{ij}^{P, A}$ represents the intra (for $i=j$) and interspecific (for $i \neq j$)  competition between plant or pollinators.  Usually $\beta_{ii} >> \beta_{ij}$ and so we assumed $\beta_{ii} = 1$ and $\beta_{ij} = 0$, for all plant and pollinators. $m^{P}_{ij} (=\frac{\gamma^{P}_{ij}A_j} { 1+h\sum_{j=1}^{S_A}{\gamma^{P}_{ij}A_j}})$ is the per-capita mutualistic benefit received by  plant $i$ from the pollinator $j$ and similarly $m^{A}_{ij} (=\frac{\gamma^{A}_{ij}P_j} { 1+h\sum_{j=1}^{S_P}{\gamma^{A}_{ij}P_j}})$ is the per-capita mutualistic benefit received by  pollinator $i$ from the plant $j$. The parameters $\gamma_{jj}^{P}$ and $\gamma_{jj}^{A}$ are the strength of mutualistic interactions, which takes the following form $\gamma_{ij} = \delta_{ij} \frac{\gamma}{d_{i}^{\rho}} $. Here $\delta_{ij}$'s  are the elements of the adjacency matrix of the network,  $\delta_{ij} = 1$ if plant $i$ and pollinator $j$ is connected, and    $\delta_{ij} = 0$ otherwise. $\gamma$ is the normalized mutualistic strength and $d_{i}$ is the degree if the $i-th$ plant or pollinators. Here the parameter $\rho$ determines the trade-off between mutualistic strength and the degree of the species and hence is associates mutualism with the network topology.  $\rho = 0$ means the mutualistic strengths are independent of the network structure. In contrast, $\rho = 1$ means there is a full trade-off, gain from the mutualism of a species from the interacting species is splitted by the number of interactions and weakened the mutualism between each interacting species. Between the two extreme cases, we took $\rho = 0.5$ following previous studies \citep{rohr2014structural, jiang2019harnessing, meng2020tipping}. $h$ is the half saturation constant, as the mutualistic benefit will saturate with the abundance of the interactive partners, and the Holling type response was first introduced by \cite{okuyama2008network} in mutualistic network model. $\kappa_{i}^{A}$ is the decline rate of the pollinators due to the external effects, and we took $\kappa_{i}^{A} = \kappa$ for simplicity. Finally, $u^{P}$ ($u^{A}$) are the constant immigration rates of plant (pollinators), which takes typically small value and thus have a little effect on the dynamics.

\subsubsection{Incorporation of the effect of contamination}
\label{incorporation}

A simple approach to model the internal concentration of contamination (or body burden) of a species, i.e., the ratio of the total contamination accumulation to the biomass, can be written by the following form \citep{luoma2005metal,veltman2008cadmium}:
 \begin{equation}
     C_{I}^{'}=(\ I_{E}+I_{F} )\ -k_{e}C_{I}
     \label{general body burden rate change}
 \end{equation}
 where $C_{I}$ is the internal concentration of the contamination time $t$. $I_{E}$ and $I_{F}$ are the inflow of the contamination coming from environment and food, respectively, and $k_{e}$ is the constant loss rate. A general assumption in modeling the dynamics of a system under the effect of contamination is that, the dynamics of the internal contamination concentration is much faster compared to the  population dynamics. So it approaches the steady state before the a significant change in the population dynamics \citep{huang2015impact,prosnier2015modeling,kooi2008sublethal}. By equating $C_{I}^{'}$ to $0$ in equation \ref{general body burden rate change}, we get the steady-state approximation of the internal contamination concentration ($C_{ISS}$) as:
\begin{equation}
    C_{Iss}=\frac{I_E + I_F}{k_{e}}, 
\end{equation}
which is basically proportional to the sum of uptakes from the environment and food. We can write $I_{E}$, $I_{F}$ as,
 \begin{equation}
\label{IE and IF forms}
\begin{split}
&I_{E}=k_{u}\times C\\
&I_{F}=A\times E\times C_{F}
\end{split}
\end{equation}
 $C$ is the concentration of the contamination in the environment, $k_{u}$ denotes uptake constant in the body,  A is assimilation efficiency, E is egnation rate, $C_{F}$ is the concentration in the food intake.

\begin{figure}[H]
\begin{center}
        \includegraphics[width=0.8\textwidth]{ 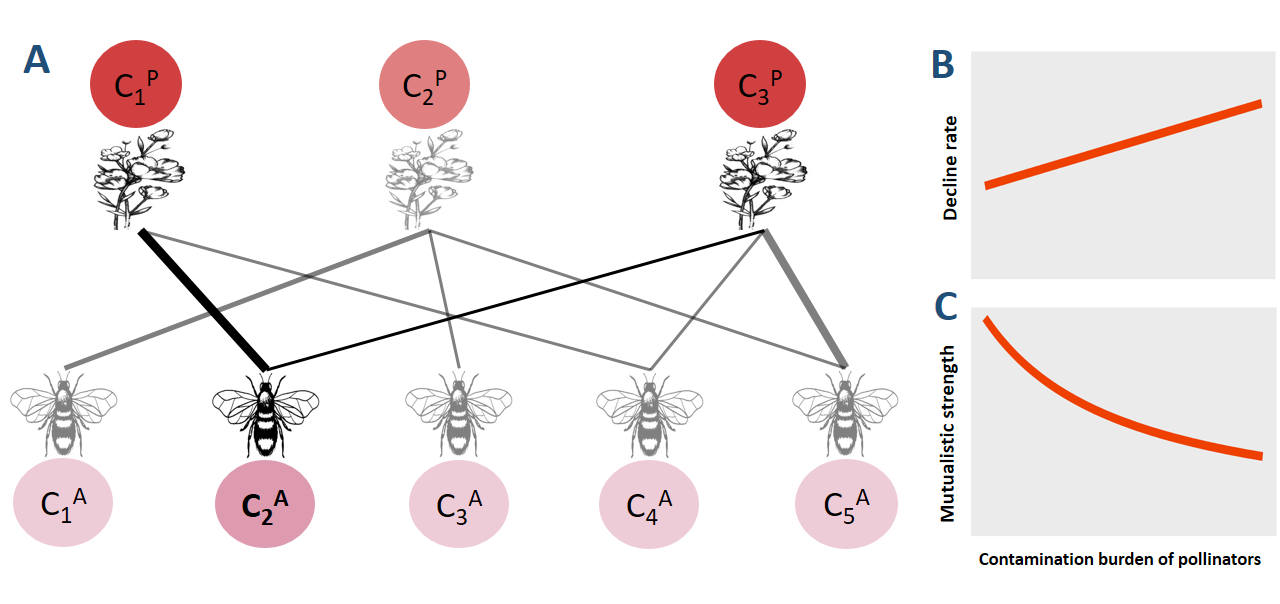}
    \end{center}
    \caption{
     Schematic representation of the  contamination model. A mutualistic network with 3 plant and 5 pollinators is illustrated, where the width of the links indicates the mutualistic strength. Contamination body burdens of plants and pollinators are indicated by the circles with the species.   As example, pollinator 2 interacts with the 1st and 3rd plant species, so the contamination burden of the pollinator will depend on the contamination burdens of the plants together with their mutualistic strength. 
    }
    \label{schematic model}
\end{figure}

Plants receive contamination directly due to the use of pesticides and other agrochemicals. So $I_F = 0$ in case of plants. Let $k_n^{i}$ be the net uptake rate of the $i-th$ plant. Then using equation \ref{IE and IF forms}, the contamination burden of the $i-th$ plant, $C_i^{P}$ becomes
\begin{equation}
    C_i^{P} = k_n^{i}\times C_i,
    \label{plant contamination burden}
\end{equation}
where $C_i$ is the contamination concentration imposed into the plant.  Now pollinators accumulate contamination, only when they visit plants and there is no source of direct contamination intake from the environment. Thus $I_E = 0$ for pollinators.  The contamination body burden of pollinator $i$, $C_{i}^{A}$,  depends on the contamination burden of the plant $j$ it interacts with (i.e. $C_{j}^{P}$) together with the mutualistic benefit it receives from that plant, $m_{ij}^{A}$, and net assimilation coefficient, $s_i$ (see Fig. \ref{schematic model}.A). Combining the terms according to equation \ref{IE and IF forms} and taking the summation over all plants, the expression of $C^{A}_i$ becomes 
\begin{equation}
    C^{A}_i = s_i \sum_{j=1}^{S_{P}} m^{A}_{ij} C^{P}_j.
    \label{pollinator contamination burden}
\end{equation}
 For simplicity, we assumed $k_n^{i} = k_n$, $C_i^P = C$ for all plants $i$ and $s_i = s$, for all pollinator $i$. 

{\it \bf Responses due to the contamination:} Commonly used pesticides and insecticide mainly contains a chemical named Neonicotinoid, which is accumulated by the pollinators and has variety of lethal and sublethal effect on the pollinator growth. Plenty of experimental studies found the positive association between the contamination burden of the pollinator with their direct mortality (see \cite{henry2012common, whitehorn2012neonicotinoid, kasiotis2014pesticide}). So we assumed that contamination mediated decline rate of the pollinators, $\hat{\kappa_{i}}$,  as a linearly increasing function of its body burden;
\begin{equation}
    \hat{\kappa_i} = \kappa_i + \sigma_i^{\kappa} C^{A}_i,
    \label{response in death}
\end{equation}

where $\sigma_i^{\kappa}$ is the effect parameter (see Fig. \ref{schematic model}.B). Furthermore, the pollinator visitation rate in plants decreases with the accumulated toxin which in turn decreases the mutualistic strength between plants and pollinators \citep{sponsler2019pesticides, wu2011sub, schneider2012rfid, piiroinen2016chronic}. Keeping this in mind, our modified mutualistic interaction strengths  $\gamma^{A}_{ij}$ (of pollinator $i$ with plant $j$) and $\gamma^{P}_{ij}$ (of plant $i$ with pollinator $j$) takes the form: 
\begin{equation}
\begin{split}
     &\hat{\gamma^{A}_{ij}} = \frac{\gamma^{A}_{ij}}{1+ \sigma_i^{\gamma_A} C^{A}_i},\\ 
     &\hat{\gamma^{P}_{ij}} = \frac{\gamma^{P}_{ij}}{1+ \sigma_i^{\gamma_P} C^{A}_j},
\end{split}
    \label{response in mutualism}
\end{equation}
which is a decreasing function of contamination body burden of pollinator $i$ (Fig. \ref{schematic model}.C). We assumed $\sigma_i^{\kappa}=\sigma^{\kappa}$ and $\sigma_i^{\gamma_A}=\sigma^{\gamma_A}$, $\sigma_i^{\gamma_P}=\sigma^{\gamma_P}$ for all pollinators and plants. Finally, pesticides can increase the plant growth rate by killing the harmful pests. So the plant growth rate takes the form: 

\begin{equation}
    \hat{\alpha_i^P} = \alpha_i^P +  \frac{\zeta_i C^{P}_i}{1+ C^{P}_i},
    \label{response in plant growth}
\end{equation}
which is a monotonically increasing function of the pesticide used and saturates at some specific value.

{\it \bf Final model:} 
Incorporating the responses of the contamination from equation \ref{response in death}, \ref{response in mutualism} and \ref{response in plant growth}, our final model becomes: 

\begin{equation}
\label{network model without C}
\begin{split}
&\frac{dP_i}{dt}=P_i\bigg(\hat{\alpha^{P}_{i}} -\sum_{j=1}^{SP}{\beta^{P}_{ij}P_j}+   \frac{ \sum_{j=1}^{SA} \hat\gamma^{P}_{ij}A_j} { 1+h\sum_{j=1}^{SA}{\hat\gamma^{P}_{ij}A_j}} \bigg) +u^{P}\\
&\frac{dA_i}{dt}=A_i\bigg(\alpha^{A}_{i}-\sum_{j=1}^{SA}{\beta^{A}_{ij}A_j}+   \frac{ \sum_{j=1}^{SP} \hat{\gamma^{A}_{ij}} P_j} { 1+h\sum_{j=1}^{SP}{ \hat{\gamma^{A}_{ij}} P_j}} \bigg)- \hat{\kappa^{A}} A_{i} +u^{A}
\end{split}
\end{equation}

\subsection{Simulations to meet our objectives}
\label{simulations}

\subsubsection{Objective 1}
\label{obj1}
To meet our first objective, i.e., to study the role of toxin on the persistence and abrupt collapse of the community, we choose $\gamma$ as one of the bifurcation parameter.  Mutualistic communities experiences abrupt and irreversible collapse with deteriorating mutualistic strength, i.e., there exists a certain threshold of $\gamma$, below which whole community collapses. To study the role of contamination on the threshold of $\gamma$, we plotted equilibrium abundances of the system with respect to the bifurcation parameter $\gamma$, for systematically increasing contamination level. 

Further to study the role of contamination on the abrupt collapse of the system, we plotted the equilibrium abundance of the system with respect to contamination level $C$ as a bifurcation parameter, for different levels of mutualistic strength ($\gamma$).   

We first demonstrate the above results using four empirical networks: A ($S_{A} = 61, S_{P} = 17$ and the number of links $L= 146$) from empirical data from Hicking, Norfolf, UK; B ($S_{A} = 38, S_{P} = 11$ and $L= 106$) from Tenerife, Canary Islands; C ($S_{A} = 44, S_{P} = 13$ and $L= 143$) from North Carolina, USA; D ($S_{A} = 42, S_{P} = 8$ and $L= 79$) from Hestehaven, Denmark (available in Web Of Life database (\url{https://www.web-of-life.es}), with ID 6, 8, 25 and 38, respectively).

Further to illustrate the generality of the result, we plotted the minimum mutualistic strength levels (below which system collapses) for increasing levels of contamination and also maximum contamination tolerance level (after which system collapses) for decreasing mutualistic strength for all 105 empirical networks, described below. 

{\it \bf Description of the data set:} We used 105 real mutualistic networks in our study from the available 153 networks in the Web Of Life database (\url{https://www.web-of-life.es}). We first excluded the very large networks which contains 100 or more species for the sake of simulation run time. Next we excluded that which contains more plants than pollinators, as this is a basic structural properties of all empirical mutualistic networks (ref...).   These two filtration leads to ultimately 105 networks with variety of network architectural properties and covers wide range of geographic locations.

\subsubsection{Objective 2}
\label{obj2}
Our second objective in this study was to verify the role of different network architectural properties on maintaining the species coexistence. We mainly focus on four network properties: (i) {\it \bf species richness}, which is the total number of surviving species in the community; (ii) {\it \bf connectence or linkage density}, the proportion of the realised link to the total number of possible links in the community; (iii) {\it \bf nestedness}, the measure of the tendency for nodes to interact with subsets of the interaction partners of better-connected nodes \citep{bascompte2003nested, bascompte2007plant, bastolla2009architecture}; (iv) {\it \bf modularity}, the measure of the tendency of the network to subdivide into modules where species belonging to the same module interacts more than other modules \citep{olesen2007modularity, jordano1987patterns}. We calculated nestedness using the formula

\begin{equation}
    N= \frac{  \sum_{i<j}^{SP} N_{ij} + \sum_{i<j}^{SA} N_{ij}       }  { \frac{S_{P}(S_{p} -1)}{2} + \frac{S_{A}(S_{A} -1)}{2}  },
\end{equation}

 following \cite{bastolla2009architecture}. Here $S_{A}$ and $S_{P}$ are the pollinator and plant richness, $N_{ij} = \frac{n_{ij}}{min ( n_{i}, n_{j})}$, $n_{i}$ is the degree of species $i$ and $n_{ij}$ is the number of times species $i$ and $j$ interacts with the same mutualistic partner. Further we calculate nestedness using the modularity function using "igraph" package in R software. We find the level of minimum mutualistic strength  for varying contamination level and also the maximum toxin tolerance for different level of mutualistic strength, for all the 105 empirical networks. We finally find the correlation between the threshold of the tipping parameters with the four network properties to find their role in species coexistence under the influence of contamination in the environment.

Plant-pollinator mutualistic networks are highly nested compared to antagonistic networks \citep{bascompte2003nested, bascompte2013mutualistic, burgos2007nestedness, bastolla2009architecture}, that is, interaction partners of specialist species are proper subsets of that of the generalist species. Pollinators, which interacts with common plants, have indirect positive effect on each other. Nestedness promotes this indirect positive effect by creating a central core of interactions, and thus increase biodiversity \citep{bascompte2003nested, bastolla2009architecture}. But the dependency on each other increases with abundance loss and as a result, the whole community collapses together abruptly when a tipping point is reaches, due to its nested structure \citep{lever2014sudden}.   To investigate the role of nestedness on the species coexistence and the tipping of the community under the influence of contamination, we find the minimum mutualistic strength threshold and maximum contamination tolerance levels, for varying the nestedness of a network. To do so, we generated mutualistic networks with specified connectence and nestedness using the algorithm proposed in \cite{medan2007analysis}, further used by \cite{lever2014sudden}. Results were averaged over 10 replications. We state the algorithm to generate a mutualistic network with specified nestedness in brief below.

{\it \bf Simulated network:} We used the algorithm proposed by \cite{medan2007analysis}, further used by \cite{lever2014sudden}, to generate a mutualistic network with specified nestedness. First, connectence and forbidden links are fixed. Forbidden links are interactions that cannot occur within a community due to morphological or phenological uncoupling (see \cite{jordano2003invariant}). A network with random structure is formed with the given connectence and forbidden links. We excluded the networks with isolated species.  Interaction between species are rearranged in a 'rich gets richer' mechanism to generate a network with desired nestedness. In each iteration, two interacting species, $x$ and $y$ is chosen randomly. Now the link between $x$ and $y$ is removed and $x$ interacts with a new randomly selected species $z$, if $z$ has more degree than $y$. This iteration process keeps the connectence unchanged and increases nestedness. Iteration stops when the desired nestedness is obtained. This algorithm enables us to vary nestedness in a mutualistic network with fixed species richness and connectence.  

\subsubsection{Objective 3}
\label{obj3}
We finally proposed a possible intervention strategy by choosing a target plant and then applying the intervention. The detail of the intervention and simulation process is given in section \ref{Intervention strategy}.

\section{Results}
\label{results}

  \subsection{Effect of contamination on the persistence of the system}
  \label{Effect of contamination on the persistence of the system}
  
\begin{figure}[H]
\begin{center}
        \includegraphics[width=1 \textwidth]{ 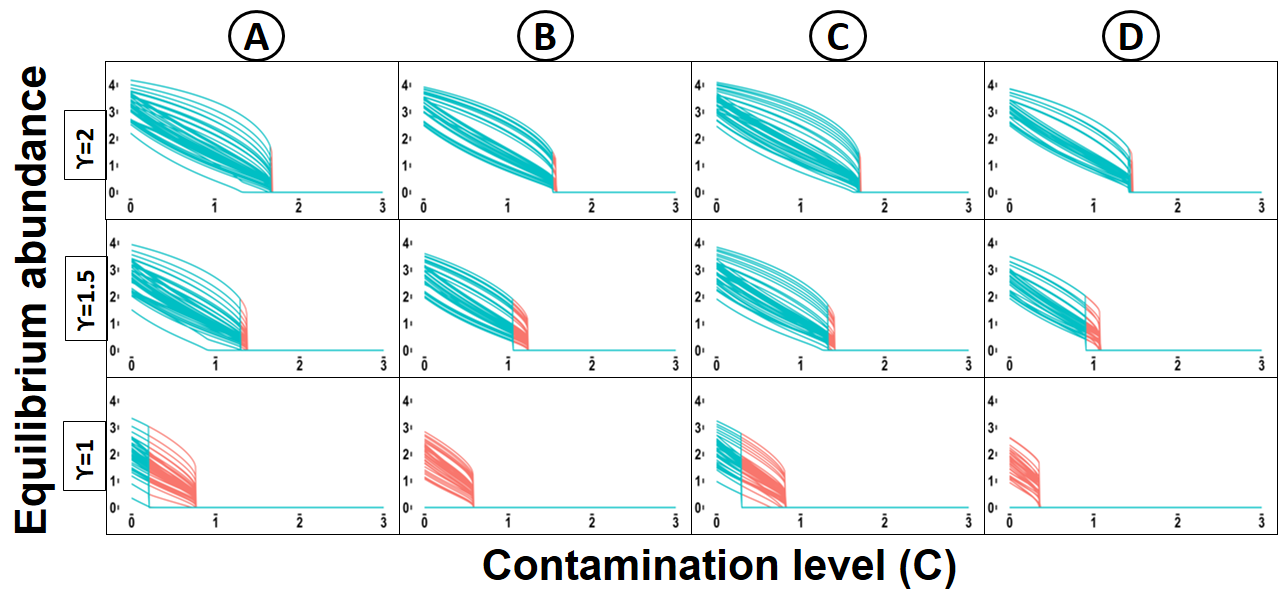}
    \end{center}
    \caption{ Equilibrium abundance of the species for the networks \textcircled{A}- \textcircled{D} with respect to increasing contamination. Abundance decreases with $C$ and contamination mediated abrupt community collapse is seen for any level of mutualistic strength. Red curves indicate the collapse and green represents recovery. MCT decreases with decreasing $\gamma$. Here $\alpha_{A}=\alpha_{P}=-0.3$, $\mu_{A}=\mu_{B}=0.0001$,
     $h=0.2$, $\kappa_{A}=0.1$, $\rho=0.5$, $\sigma^{\kappa}=0.1$, $\sigma^{\gamma A}=0.5$, $\sigma^{\gamma P}=0.1$, $\zeta = 0.1$. }
    \label{codim1_C_varygamma}
\end{figure}

\begin{figure}[H]
    \begin{center}
        \includegraphics[width=1 \textwidth]{ 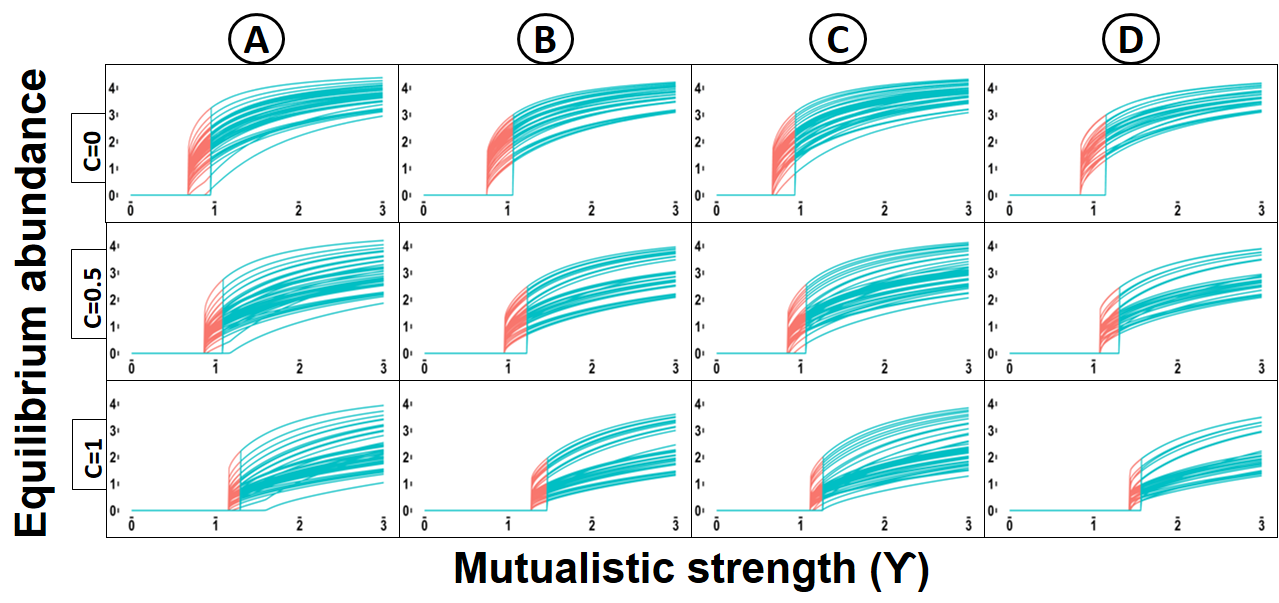}
    \end{center}
    \caption{ Threshold level of $\gamma$ at which community collapses (MGC), increases with contamination level ($C$). Parameters are same as in Fig. \ref{codim1_C_varygamma}. }
    \label{codim1_gamma_varyC}
\end{figure}

In this section, we demonstrate the effect of contamination on the coexistence and abrupt transition of a mutualistic community. We first used 4 empirical networks, \textcircled A-\textcircled D (described in Section. \ref{obj1}), to demonstrate the results (in Fig. \ref{codim1_C_varygamma}, and \ref{codim1_gamma_varyC}). As we increase the contamination level, the abundance of each species in the community decreases. This is expected, as increasing contamination in plants increases the contamination body burden of the pollinators, which in turn increases their mortality and decreases the mutualistic benefit received from plants. These effects are responsible for the reduction in abundance.   Surprisingly, with gradual increase in the contamination level, the system experiences a catastrophic transition from the stable coexistence state to an alternative state of very low abundance (see Fig. \ref{codim1_C_varygamma}). The orange lines represents the collapse and green lines represents the recovery of the system. This phenomenon of sudden changing of a system's state in response to some system parameters is known as tipping \citep{lever2014sudden, jiang2018predicting}. Contamination triggers the two tipping parameters of the mutualistic system, decline rate of the pollinators and the strength of the mutualism. As a results, abrupt and irreversible collapse of the population is seen in the community with increasing contamination. The threshold contamination level at which the tipping occurs, is denoted as maximum contamination tolerance (hereafter MCT). We see that, as the average mutualistic strength ($\gamma$) of the community decreases, MCT decreases (Fig. \ref{codim1_C_varygamma}). That is the systems with low average mutualistic strength has low potential to coexists in a contaminated environment.    

After the immediate collapse of the community, decreasing contamination level does not bring back the system to its previous state. System recovers at relatively low level of contamination, thus forming a hysteresis loop. At this parameter window, there is a bistability in the system and so the initial abundance determine its persistence. The parameter window for bistability increases with decreasing $\gamma$, which means that even before the whole community collapse, the system is sensitive to its initial abundance. For the community with very low mutualistic strength ($\gamma$), the MCT is comparably very low. Furthermore, the region of bistability is broad and sometimes there is no recovery from the collapse (see Fig. \ref{codim1_C_varygamma}, for $\gamma = 1$). 

To understand the influence of contamination on the tipping of a community, we plot the equilibrium abundance with respect to $\gamma$, for different level of contamination level in the environment (see Fig. \ref{codim1_gamma_varyC}). Abundance decreases with decreasing $\gamma$, and then experiences an abrupt collapse at a threshold, below which the system is in the alternative stable state of very low abundance. We call it minimum gamma for coexistence, hereafter MGC.   We see that MGC increases significantly with increasing contamination, for all the 4 networks. Also the threshold of the recovery, after which system backs to its previous stable coexistence state irrespective of all initial conditions, increases with increasing contamination. As example, network A coexists, irrespective of initial condition, for the mutualistic strength $\gamma = 1$, when there is no contamination in the system. But the system posses bistability at  $\gamma = 1$ for $C=0.5$ and even extinction state for $C=1$ for any initial abundance.

\begin{figure}[H]
\begin{center}
        \includegraphics[width=1 \textwidth]{ 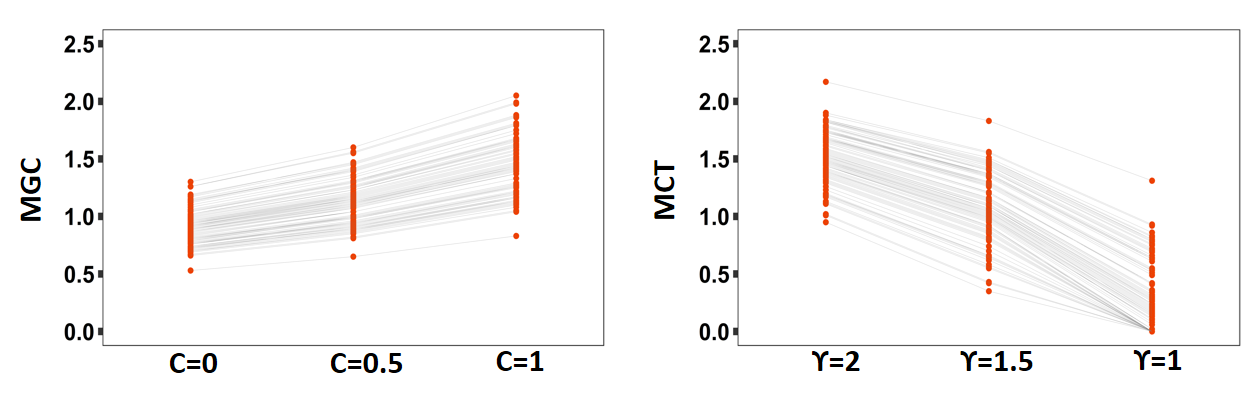}
    \end{center}
    \caption{ MGC's and MCT's for all 105 empirical networks. (a) MGC increases with increasing contamination level, (b) MCT decreases with decreasing average mutualistic strength $(\gamma)$. Parameters are same as in Fig. \ref{codim1_C_varygamma}.}
    \label{MGC_MCT_all}
\end{figure}

Further, to verify the prevalence of our findings on the effect of contamination on the coexistence and tipping of a mutualistic community, we plot the MCT for different level of $\gamma$ and MGC for varying contamination level ($C$), for all 105 empirical networks mentioned in section \ref{obj1} (Fig. \ref{MGC_MCT_all}). For all the networks, the MCT decreases significantly with decreasing $\gamma$ (Fig. \ref{MGC_MCT_all}.A). Also MGC increases with increasing contamination (Fig. \ref{MGC_MCT_all}.B). Which implies that in the presence of contamination, plant-pollinator communities with high average mutualistic strength ($\gamma$) will coexists. Also the communities with low $\gamma$ is prone to catastrophic collapse in the face of contamination.

\subsection{The role of network topological properties in maintaining the coexistence}
\label{The role of network topological properties in maintaining the coexistence}

\begin{figure}[H]

    \begin{center}
        \includegraphics[width= \textwidth]{ 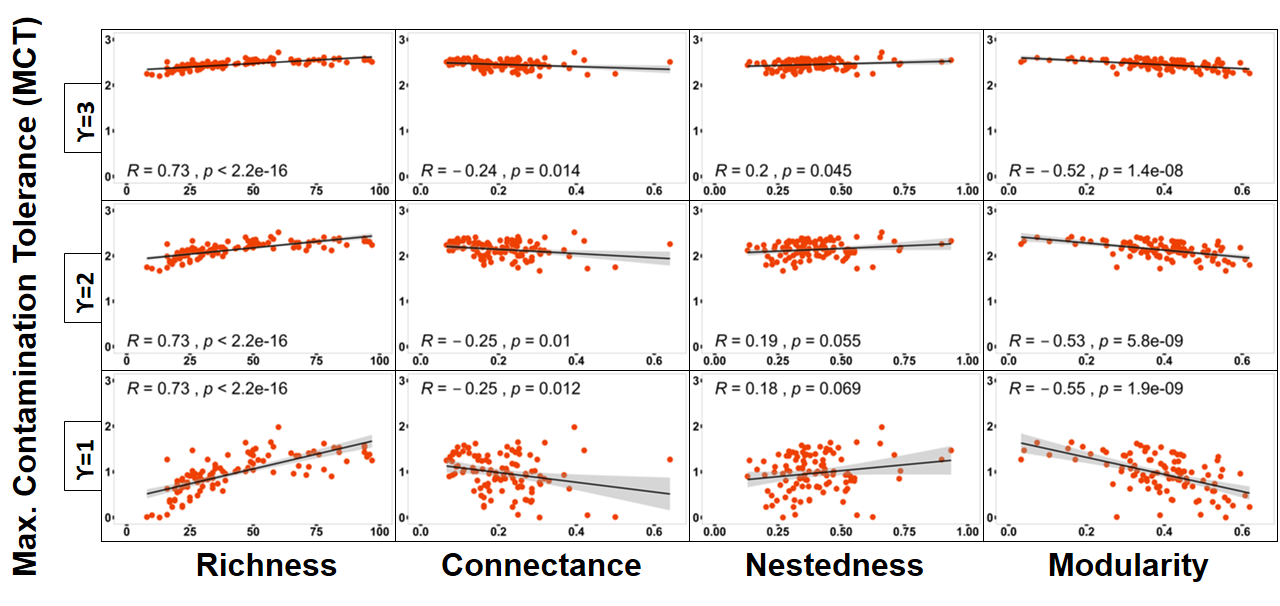}
    \end{center}
    \caption{ Pearson correlation coefficients of the MCT's for all 105 networks, with the four network architectural properties. Richness and nestedness is positively correlated with MCT, whereas connectence and modularity shows the opposite trend. Here $\alpha_{A}=\alpha_{P}=-0.1$, other parameters are same as in Fig. \ref{codim1_C_varygamma}. }
    \label{MCT_all}
\end{figure}

\begin{figure}[H]

    \begin{center}
        \includegraphics[width= \textwidth]{ 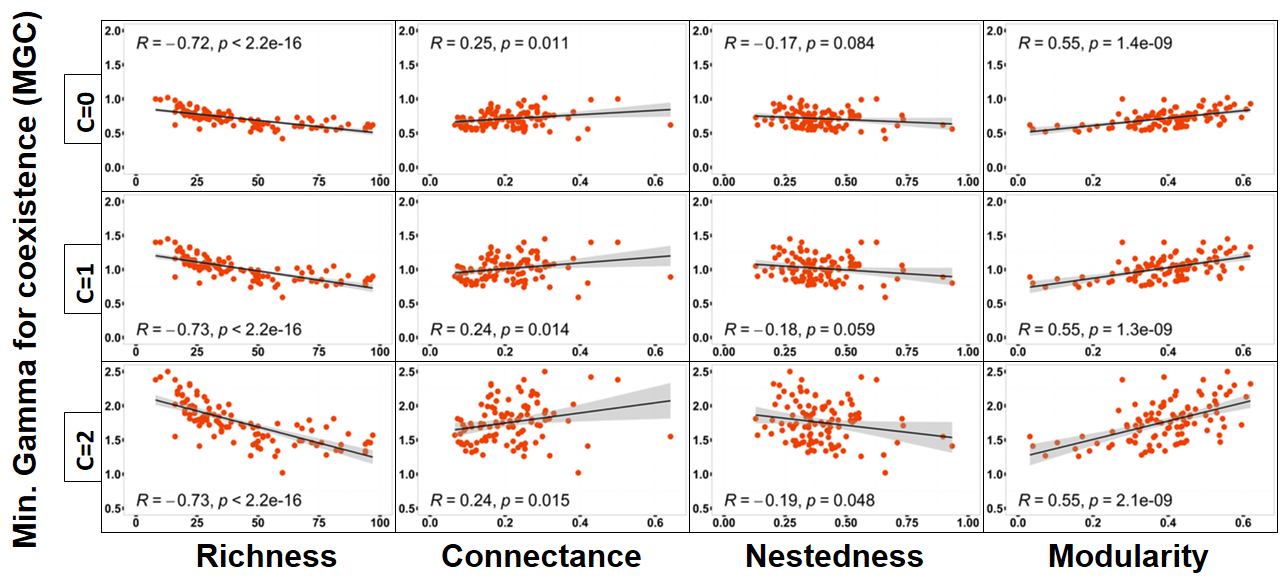}
    \end{center}
    \caption{Pearson correlation coefficients of the MGC's for all 105 networks, with the four network architectural properties. Richness and nestedness is negatively correlated with MGC, whereas connectence and modularity shows the opposite trend. All parameters are same as in Fig. \ref{MCT_all}. }
    \label{MGC_all}
\end{figure}

In this section, we study the role of several network architectural properties in maintaining the coexistence and the tipping of the community in the face of contamination.    To do so, we plot MCT for all 105 networks with four important properties of a mutualistic network, richness, connectence, nestedness and modularity mentioned earlier, with varying average mutualistic strength $\gamma$  (Fig. \ref{MCT_all}). In order to understand the linear dependency of the MCT of a community with the network properties, we mentioned the Pearson correlation coefficient between them in each plot. The same analysis is done for the case of MGC with varying contamination level in Fig. \ref{MGC_all}.

Species richness or diversity of a community has a significant positive correlation with the MCT, which means species rich communities can delay the abrupt collapse due to contamination (Fig. \ref{MCT_all}, 1st column). As $\gamma$ increases, MCT increases for all the networks, but surprisingly, correlation remains the same. Also richness is negatively correlated with MGC, for any contamination level (Fig. \ref{MGC_all}, 1st column). This implies networks with higher richness can coexists with minimal average mutualistic strength $\gamma$ under the toxin.  Thus richness is beneficial for the persistence of a mutualistic community in the presence of contamination, which is agreeable to the previous studies where community size increases resilience of a mutualistic community \citep{okuyama2008network} and also for the food webs \citep{garay2013more}. Connectence or linkage density has a positive correlation with the MGC, for varied level of contamination (Fig. \ref{MGC_all}, 2nd column). Also connectence decreases the ability of a network to persists under contaminated environment as it negative correlated with the MCT (Fig. \ref{MCT_all}, 2nd column). So a densely connected mutualistic community is found to be more prone to extinction in the presence of contamination, which is opposite for the case of food web under pollution, where connectence is positively correlated with the persistence of the system \citep{garay2013more}. Pollinators accumulate more contamination in a densely connected network. So their body burden increases, which acts antagonistically.  Nestedness is beneficial for a mutualistic community as it is positively correlated with MCT (Fig. \ref{MCT_all}, 3rd column) and negatively correlated with MGC  (see Fig. \ref{MCT_all}, \ref{MGC_all}, 3rd column, respectively). This indicates that nested communities is more robust in the face of contamination in the environment. Species in a nested networks are adhesively connected to a central core of interaction and creates a  positive feedback loop between the interacting species. Harsh condition increases the dependency between the interacting species and this is why nestedness is beneficial for a community under environmental stress. The result is synergistic with previous studies \citep{saavedra2013estimating, rohr2014structural}, where nestedness is shown to increase the tolerance level of a community in a fluctuating environment. Modularity acts opposite as nestedness  as it has strong negative (positive) correlation with MCT (MGC) ( Fig. \ref{MCT_all}, \ref{MGC_all}, 4th column, respectively). A modular network is prone to collapse under environmental stress, which is opposite of the antagonistic interactions \citep{garay2014food}. As species interactions are more restricted to specific modules, pollinators are more dependent with the plants within modules, the biomass loss of which can promote the collapse of that specific module, which in turn can trigger the extinction of the whole community.  

Overall MCT decreases with decreasing mutualistic strength and MGC increases with increasing contamination, for all 105 real networks studied. Large mutualistic community with high nestedness, low connectence and low modularity is found to be capable to functioning in the face of environmental contamination. Correlation between the network metrices and MCT (MGC) remains unchanged with varying the level of mutualistic strength (contamination). Which means that their linear associationship remains unchanged over varying conditions. But the slope of the regression line increases with increasing contamination level or decreasing mutualistic strength, in all the cases.  So the community is more dependent with its topological properties. Also the variance of the data is higher, i.e., data points more scattered in the extreme conditions.  This implies the tipping points of the community collapse largely varies over networks in harsh condition (i.e., less mutualistic strength and higher contamination level).

\begin{figure}[H]
\label{nestedness_C}
    \begin{center}
        \includegraphics[width=0.8\textwidth]{ 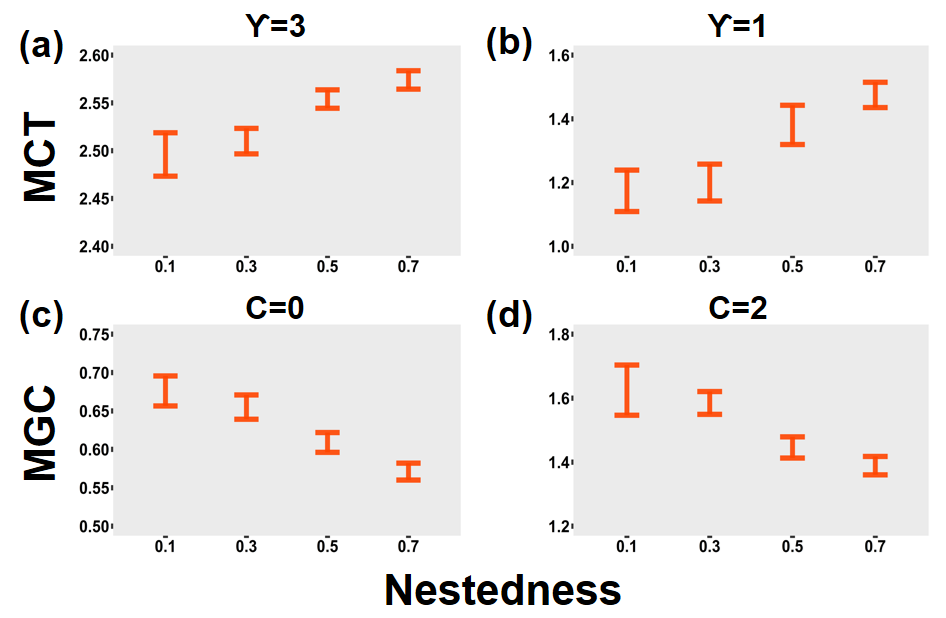}
    \end{center}
    \caption{ Mean + - sd of the MGC and MCT's for simulated mutualistic communities with increasing nestedness. Here number of plants, $S_{P}=10$, number of pollinators, $S_{A}=40$, and  $C=0.2$, $Fl=0.2$. MCT (MGC) increases (decreases) with nestedness, for varying level of mutualistic strength and contamination level. Results are replicated 10 times. Other parameters are same as in Fig. \ref{MCT_all}. }
     \label{nestedness1}
\end{figure}

Mutualistic networks are often nested, entailing a core of species with many interactions among themselves, species with few interactions interacting with proper subsets of species with many interactions and few if any interactions among species with few interactions. Nestedness influences the positive stability-complexity relationship of a mutualistic community by increasing the positive feedback loop \citep{okuyama2008network}. Nestedness reduces effecting interspecific competition \citep{bastolla2009architecture} and increase biodiversity. Nestedness increases the community resilience to species extinction \citep{memmott2004tolerance, burgos2007nestedness}.  To verify the role of nestedness on the potential of the community to coexist under the contaminated environment, we plot MCT and MGC for simulated mutualistic networks with increasing nestedness (see Fig. \ref{nestedness1}). We use the algorithm stated in section \ref{obj2} to simulate a network with specified nestedness. The results for simulated network is synergistic with that of the real empirical networks, MCT (MGC) increases (decreases) with nestedness. Also the positive effect of nestedness on evading the tipping significant under the deteriorating environment.  Overall the more nested a network, the more ability to functioning properly under the effect of contamination.

\section{Intervention strategy}
\label{Intervention strategy}

In this section, we developed an ecologically feasible strategy to manage or delay the  critical transition of a mutualistic community.  Managing critical transitions or tipping typically means delaying the global extinction or altering global extinction to gradual extinction of individual species. \cite{jiang2019harnessing} proposed a biologically feasible intervention strategy by selecting a targeted pollinator species and then controlling its abundance or fixing its decay rate. This simple but effective intervention strategy can mitigate the abrupt collapse of a plant-pollinator community.  In our system, we dealt with the agrochemical use in plants which is contaminant to the pollinators. So we propose an intervention strategy related to the management of the use of agrochemicals viz. pesticides, herbicides etc. in plants.  A recent field study conducted by \cite{pecenka2021ipm} claimed that decrease of the percentage of pesticide use can dramatically increase the crop production. Inspired from this, we make our strategy by selecting a "targeted plant" from the community by a systematic way and then manage its contamination level, subsequent steps of which is stated below.

\begin{figure}[H]
\begin{center}
        \includegraphics[width=1\textwidth]{ 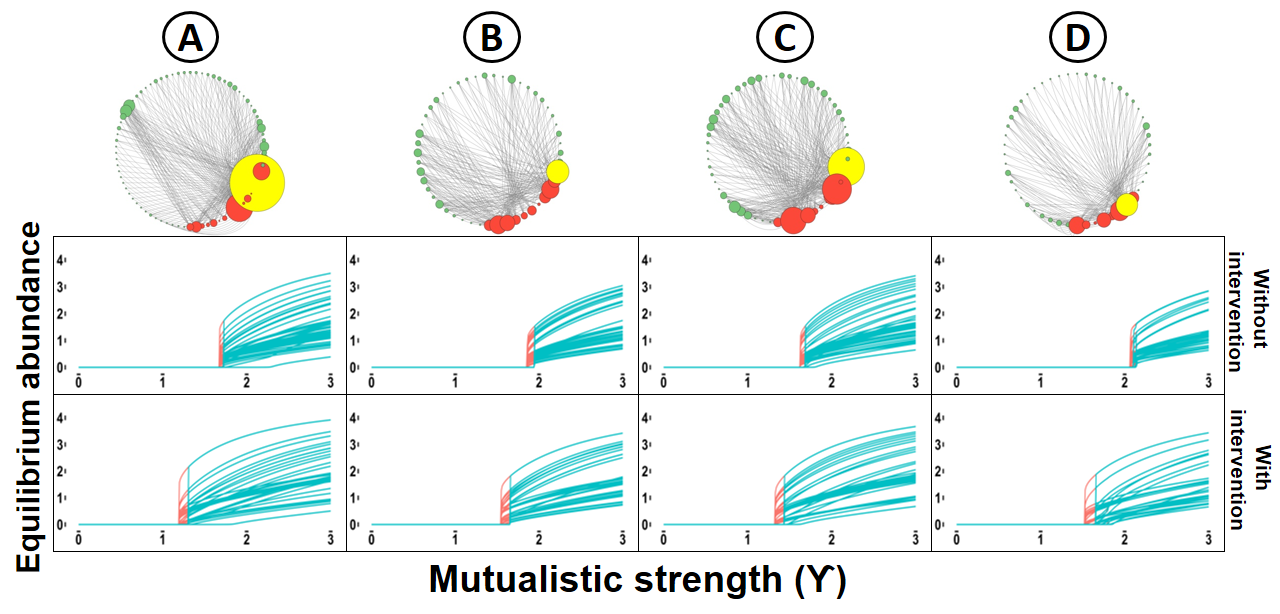}
    \end{center}
    \caption{ Equilibrium abundance of all species for the four networks \textcircled{A} - \textcircled{D}, with and without applying the proposed intervention strategy. Here green (red) circles represents the pollinators (plants) and yellow circle represents  the targeted plant. Size of the circles represents the nodal degree of the species. Parameters are same as in Fig. \ref{codim1_C_varygamma}, $C=1.5$ }
    \label{intervention_gamma}
\end{figure}

{\it \bf Target plant:} A large number of research devoted on quantifying the most important node of a network or the central node of a network and the corresponding measure is called {\it centrality}. The target plant we choose in our study is the node with highest centrality. Of course there are many alternative measures of centrality. To justify our consideration of centrality for the target plant, we state here some widely used centrality measures for bipartite networks.  Degree centrality, the number of degree a particular node have;  Eigenvector centrality, the elements of the eigenvectors for the largest eigenvalue \citep{bonacich1987power};  Page rank centrality, where a node have higher centrality if it is connected to other nodes with high centrality. We choose degree centrality over all centrality measures for its simple yet effective concept, as high degree plants acts as hubs of a plant-pollinator community. Albeit all the centrality measures gives the same nodes for most of the real networks we studied, mentioned in section \ref{obj1}.  Degree has generally been extended to the sum of weights when analysing weighted networks \citep{newman2004analysis, barrat2004architecture}. It is equal to the traditional definition of degree if the network is binary. However these two measures give same nodes for all the empirical networks we studied.

\begin{figure}[H]
    \begin{center}
        \includegraphics[width=0.6 \textwidth]{ 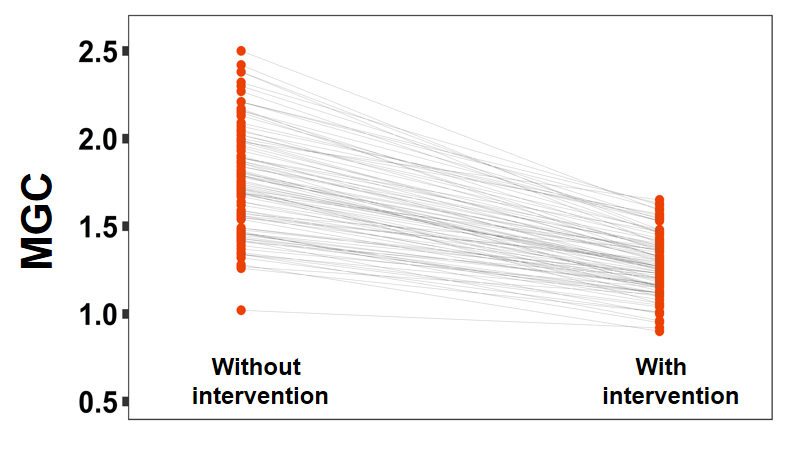}
    \end{center}
    \caption{Threshold level of the mutualistic strength below which community collapses (MGC's), for all 105 empirical networks, with and without the intervention. Targeted intervention can effectively decrease the MGC. Here $\alpha_{A}=\alpha_{P}=-0.1$, $C=2$, other parameters are same as in Fig. \ref{codim1_C_varygamma}.}
    \label{intervention_all}
\end{figure}

After choosing the target plant, we reduce the pesticide load of it by half (i.e. 50\%). First we illustrate the effectiveness of the intervention for networks \textcircled A-\textcircled D. We plot the equilibrium abundance with respect to average mutualistic strength ($\gamma$) in contaminated environment, without and with intervention strategy (see Fig. \ref{intervention_gamma}). Our results indicates that intervention strategy significantly decreases the extinction threshold of the community (i.e., MGC) with decreasing mutualistic strength. For instance, the network \textcircled A,  \textcircled C cannot survive with average mutualistic strength $\gamma = 1.5$ in the contaminated environment. But when intervention acts, the coexistence is restored, though some of the species may go to extinction before the whole community collapse, but the global extinction is delayed (see Fig. \ref{time series intervention} in Appendix \ref{Effectiveness of the intervention}, for the time series plots). Similar results holds for networks \textcircled B and \textcircled D, for $\gamma = 1.75$ and $1.8$, respectively. However extinction of few species may be regarded as a precursor of the critical transition in this case. Recovery threshold from the extinction state with increasing $\gamma$ is also shifted due to the intervention. Which means that, community recovers from the extinction state for slight increase in mutualistic strength, when intervention acts. To generalize our findings, we plot the MGC for all 105 real networks for without and with intervention (Fig. \ref{intervention_all}). We see that, intervention effectively decreases the MGC, for all real networks considered, which means that the intervention strategy we proposed has the enough potential to delay the tipping.

{\it \bf Importance of systematic targeting:} We further studied the importance of finding a targeted plant to apply the intervention strategy. To do so, we compare the effectiveness of the intervention for two cases; first, intervention in the target plant and secondly, intervention in any arbitrary plant in the community. We perform statistical significance test (see Appendix \ref{Significance of the targeted plant}) to the MGC for the two cases mentioned above and see that the MGC for the second case is significantly greater than the first case. That means the threshold deteriorating mutualistic strength level, below which community collapses, significantly decreases in case of targeted intervention. Therefore the intervention strategy is maximally beneficial when we apply it to a plant which is targeted in a systematic way.  

\section{Discussions} 
\label{discussions}

Plant-pollinator interactions are essential for terrestrial biodiversity and crop production. Chemical contamination due to excessive use of agrochemicals is an important driver of pollinator decline in the current era of anthropogenic changes as habitat loss, climate warming, parasite infection etc. But a mathematical study integrating the effect of contamination on a plant-pollinator mutualistic community is still lacking. Here we modelled a multispecies plant-pollinator community under the influence of environmental contamination, where pollinators accumulate toxin from the plants. Accumulated toxin affects pollinator decline rate and mutualistic strength between the plant and the pollinators, which in turn can act as a driver of abrupt and irreversible community collapse, from the stable coexisting state to community extinction, also known as tipping.  We studied the effect of contamination level on the persistence of the community. We also investigated the role of different network architectural properties of a mutualistic community to maintaining the coexistence of the community in the face of contamination. Finally we proposed an intervention strategy regarding maintenance of the contamination level of a single target plant which can significantly evade the catestrophic collapse of the community.

 Species abundance in a mutualistic community decreases with contamination level (Fig. \ref{codim1_C_varygamma}). Elevated pollinator decline rate and reduced mutualistic strength due to contamination causes the population decline.  Increasing contamination beyond a certain threshold can cause a catastrophic community collapse, as it triggers the two tipping elements of a plant-pollinator mutualistic community. Now a system backs to its functioning state i.e., recovers with much lower levels of contamination compared to MCT, thus forming a hysteresis loop.  Within this contamination range, the system can coexists or may go to extinction, depending on the initial population level, thus showing bistability. Contamination mediated bistability is ubiquitous in a wide range of ecological systems \citep{huang2015impact, banerjee2021chemical, chattopadhyay2022environmental} and our finding adds one extra dimension into it. The systems with lower mutualistic strength has a very low potential to function under contamination as it possess much lower level contamination tolerance (Fig. \ref{MGC_MCT_all}) and broader bistability window (Fig. \ref{codim1_C_varygamma}). Also there is sometimes no recovery from the extinction in some networks studied, that means the community is always prone to extinction, if the initial population is low. Further the MGC level below which community collapses, effectively increases with contamination for all the empirical networks studies (Fig. \ref{codim1_gamma_varyC}, \ref{MGC_MCT_all}.B), indicating that a plant-pollinator community needs higher mutualism strength to exists in the face of contamination.

Highly nested speciose communities with less connectence and modularity has the greater potential to endure contamination (Fig. \ref{MCT_all}, \ref{MGC_all}). Species richness is strongly positively correlated with MCT, the maximum threshold level of contamination after which community collapses, is relatively high for large systems.  Also MGC, the minimum mutualistic strength below which community extinct, is low for species rich communities. Therefore species diversity has positive effects on the existence of a mutualistic community under contamination stress.  Our results agrees with the conclusions of the previous studies  \citep{okuyama2008network, thebault2010stability}, where diversity is shown to be positively associated with the persistence and resilience of mutualistic communities. Connectence decreases with richness in mutualistic communities \citep{olesen2002geographic}, which is also evident from our studied empirical networks (see Fig. \ref{Correlation network properties}.A in App. \ref{Empirical networks plots}). Local stability and degree of localization (metrices of stability; ability of a system to absorb perturbations) negatively depends on the connectence of a mutualistic community \citep{allesina2012stability, suweis2015effect}. Also extinction cascades, the tendency of  secondary extinction of a mutualistic network increases with connectence \citep{vieira2015simple}. Overall connectence has negative impact on the stability of a mutualistic community (but see \cite{okuyama2008network, thebault2010stability}). Our results is synergistic with the previous findings, as the MGC (MCT) has positive (negative) correlation with the connectence. Thus densely connected networks are less likely to function under contaminated environment. Pollinator's contamination body burden increases when it is connected with more plants in a densely connected community, which in turn affects it's demographic rates. This makes the community more prone to extinction under contamination exposure. Nestedness, the anomalous property of a mutualistic network \citep{bascompte2003nested}, can boost the ability of the system to persists, especially in extreme circumstances \citep{thebault2010stability}. Species in a nested networks are cohesively connected to a central core of interaction, where generalists and specialists both interacts with generalists and specialist-specialist interactions are rare. Nestedness creates a  positive feedback loop between the interacting species and increase the diversity by reducing the interspecific competition \citep{bastolla2009architecture}. Community response to cascading extinction of pollinators is minimized for nested structure \citep{memmott2004tolerance}. Nestedness has a positive effect on community persistence, resilience and structural stability (\cite{okuyama2008network, thebault2010stability, rohr2014structural}, but see \cite{allesina2012stability, campbell2012topology}). Results from our empirical and simulated networks shows that nestedness is positively associated with system's ability to function in extreme conditions (high contamination level or low mutualistic strength). The interdependence within species in a nested community increases with deteriorating conditions which in turn delay the whole community collapse. Modularity has strong negative correlation with the system's persistence under contamination; MCT (MGC) is low (high) for the communities with high modularity. Interactions becomes restricted into modules which prevents the stabilizing mechanism of mutualistic communities; specialists to generalists interaction. That's why high modular networks has low degree of nestedness (see Fig. \ref{Correlation network properties}.B in Appendix \ref{Empirical networks plots}). Previous studies reported the negative relationship of modularity with the persistence and resilience of a mutualistic community \citep{thebault2010stability}. However nestedness may be beneficial for the stability of a food web network \citep{thebault2010stability}, especially under the contaminated environment \citep{garay2014food}, opposite to our observed trends in case of mutualistic communities.

We further proposed an ecologically plausible intervention strategy which can helps a plant-pollinator community to function upto a certain extent of the increasing environmental contamination. Our intervention strategy involves managing of the contamination level of a single plant in the community.  Reducing the contamination burden of it by half can effectively work as an intervention strategy. To apply the intervention, we choose one target plant from the community, the plant with highest interaction, i.e., highest degree. Our study indicates that managing the highest degree plant in the community can effectively evade the tipping of the community for decreasing average mutualistic strength or increasing contamination level. Pollinators, from most specialist to generalist status, are likely to interact with the highest degree plant of the network, that's why this is called the hub of the network. So reducing the contamination burden of the most connected plant or hub of the community can effectively decrease the net contamination burden of the pollinators, which implicitly reduces the detrimental effect of the contamination on pollinators. This in turn helps the community to sustain by alleviating the effect of the deteriorating environment. As a result the minimum average mutualistic strength, below which the whole community collapses, effectively decreases when intervention acts. The difference between the two sets of MGC for all 105 real network studied is statistically significant when we apply the management on any arbitrary plant in the community. So our study emphasise the importance of the highly connected plant in the case of management.  Furthermore, when intervention applies, extinction of some species is seen before the system reaches to its threshold tolerance level of the deteriorating environment,  which may be pointed out as the precursor of the whole community collapse. Also the range of the bistability parameter window, where the fate of the community is dependent on its initial population level, decreases when intervention applies to the system. A recent study by \cite{jiang2019harnessing} proposed a management strategy on plant-pollinator community by fixing the abundance or the decay rate of a specific pollinator in the system and this strategy has the potential to delay the tipping and speed up the recovery of the system. Our proposed intervention adds extra dimension in the management policy under contamination exposure, which involves management of plants.


\begin {appendices}

\section{Empirical networks plots}
\label{Empirical networks plots}

\begin{figure}[H]
\begin{center}
        \includegraphics[width=0.8 \textwidth]{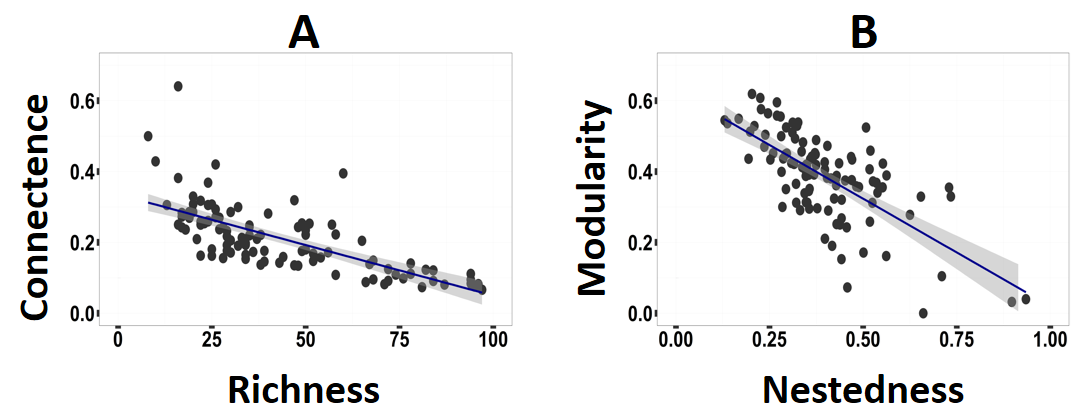}
    \end{center}
    \caption{Correlation between network architectural properties for 105 empirical networks studied. Richness-connectence and nestedness-modularity are negatively correlated. }
    \label{Correlation network properties}
\end{figure}

\section{Effectiveness of the intervention}
\label{Effectiveness of the intervention}

\begin{figure}[H]
\begin{center}
        \includegraphics[width=0.8 \textwidth]{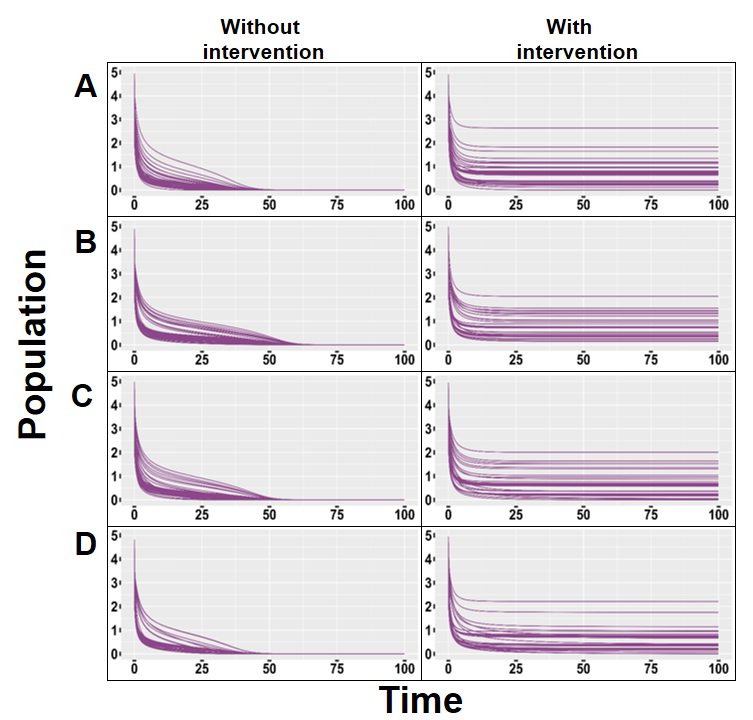}
    \end{center}
    \caption{Time series solutions for four networks \textcircled{A} - \textcircled{D}, with and without applying the proposed intervention strategy. Here $C=1.5$, and $\gamma = 1.5, 1.75, 1.5, 1.8$ for the four networks, respectively. Other parameters are same as in Fig. \ref{intervention_gamma}. Intervention can effectively restore the communities from extinction.  }
    \label{time series intervention}
\end{figure}

\section{Statistical hypothesis test for the significance of targeting a plant}
\label{Significance of the targeted plant}
We verify that intervention in targeted plants is more effective than management in arbitrary plants in the community. It is enough to prove that the MGC's for the case of randomly managed plants are significantly greater than that of targeted plants. Let $x$  and $a$ be the vector containing MGC's for the formar and latter cases, respectively. We assumed that $x$ is normally distributed with mean $m$ and let $\Bar{a}$ be the mean of $a$.   Here we are to test the hypothesis: "MGC's for random intervention is equal to MGC's for targeted intervention". 

So here we are to test the null hypothesis:

$H_0: \mu=0$, 

against the alternative hypothesis:

$H_1: \mu>0$,

where $\mu$ is the mean of the vector $y= x-\Bar{a}$, which is also a normal distribution with mean $m-\Bar{a}$ and unknown standard deviation.   

Then the suitable test statistic is given by 
$t= \frac{\sqrt{n} (\Bar{y} - 0)}{s}$, 
whose sampling distribution is a t-distribution with $n-1$ degrees of freedom, $n$ is the size of the vector $y$. $\Bar{y}$ is the mean of $y$ and $s^2= \frac{n}{n-1} S^2$, $S^2$ is the variance of $y$. 

Let us test the hypothesis $H_0$ at $5\%$ level of significance, i.e., $\epsilon=0.05$. 

Now the critical region at the significance level $\epsilon=0.05$ is ${t: t> t_\epsilon}$, where $t_\epsilon$ is given by $P(t>t_\epsilon)=\epsilon$, and $t$ has t-distribution with mean $n-1$ degrees of freedom. 

\begin{figure}[H]
\begin{center}
        \includegraphics[width=0.6 \textwidth]{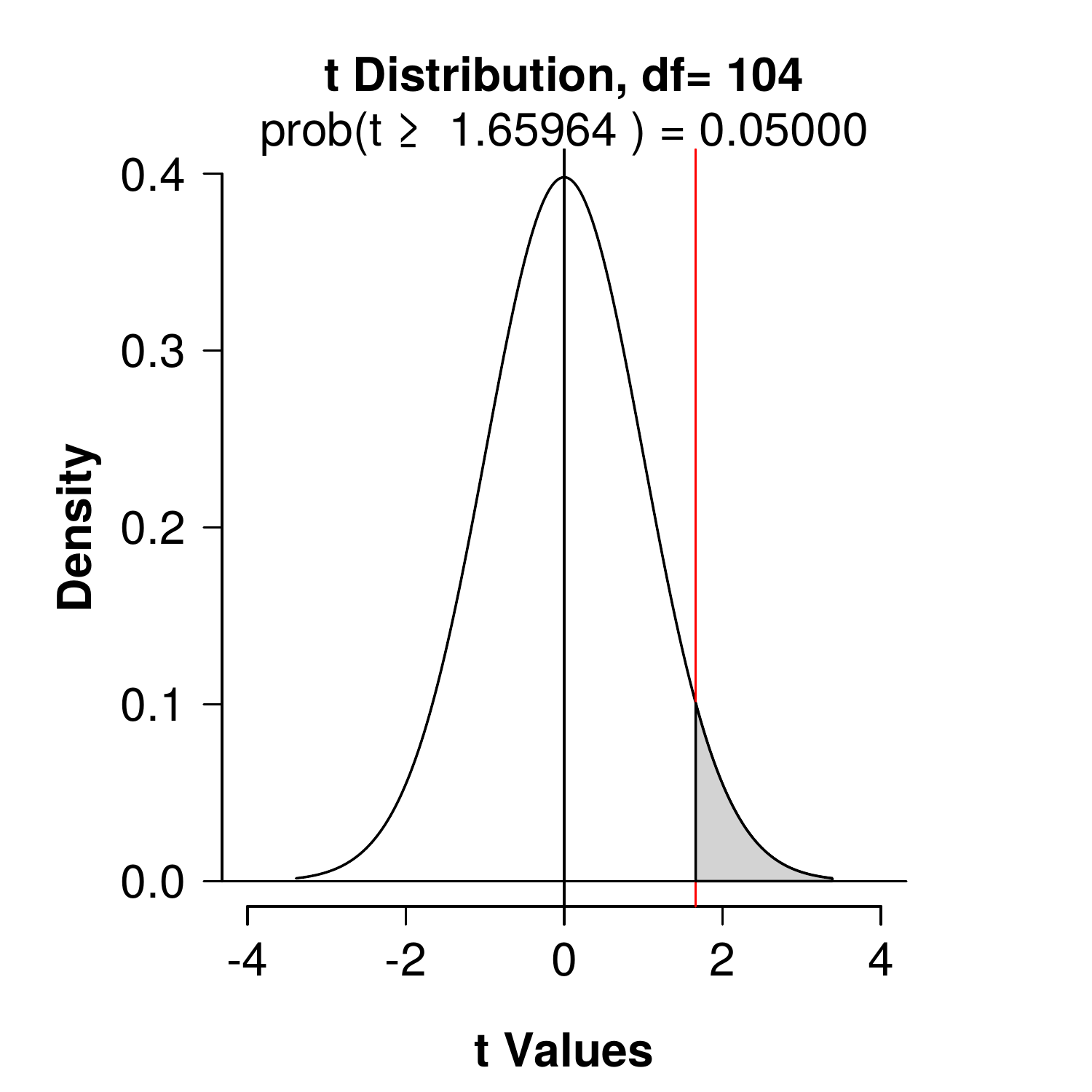}
    \end{center}
    \caption{Critical region of the for rejecting the null hypothesis is indicated by the grey area, i.e., after the red vertical line. }
    \label{t-distribution}
\end{figure}

Here $n=105$. From the density of the t-distribution with $n-1$ degrees of freedom, we have $t_\epsilon = 1.659$. Thus the critical region for rejecting the null hypothesis $H_0$ is:
${t: t> 1.659}$ (see Fig. \ref{t-distribution}).

Now from our vector $x$, the MGC's for random intervention case, the value of the test statistic is 
$t= \frac{\sqrt{n} (\Bar{y} - 0)}{s} = 10.69$, which lies in the critical region. 

Thus the null hypothesis $H_0$ is rejected and alternative hypothesis $H_1$ accepted at $5\%$ level of significance. 

So it is reasonable to believe that $\mu > 0$, i.e., MGC's for the randomly managed plants is greater that that of targeted plants. In other words, targeted intervention is more effective than intervention in arbitrary plants in the community.

\end {appendices}


\end{document}